\documentclass[12pt,a4paper]{iopart}

\usepackage[english]{babel}

\usepackage{graphicx}

\usepackage{iopams}

\usepackage{bm}

\newcommand{\text}[1]{{\rm #1}}

\usepackage{acronym}

\usepackage{hyperref}

\usepackage{cite}
\newcommand{\ii}{\rmi}
\newcommand{\ee}{\rme}
\newcommand{\dd}{\rmd}
\newcommand{\kk}{{\rm k}}
\newcommand{\jj}{{\rm j}}

\renewcommand{\Re}{{\rm Re}\,}
\renewcommand{\Im}{{\rm Im}\,}
\renewcommand{\vec}[1]{\boldsymbol{#1}}	% bold vectors

\newcommand{\eqref}[1]{(\ref{#1})}

\begin{document}

\title{Bifurcations and exceptional points in dipolar Bose-Einstein
  condensates}
\date{}
\author{Robin Gut\"ohrlein, J\"org Main, Holger Cartarius,\\ G\"unter Wunner}
\address{1. Institut f\"ur Theoretische Physik,
  Universit\"at Stuttgart, 70550 Stuttgart, Germany}
%\\[2ex]  Date: \today}

\begin{abstract}
Bose-Einstein condensates are described in a mean-field approach by
the nonlinear Gross-Pitaevskii equation and exhibit phenomena of
nonlinear dynamics.  The eigenstates can undergo bifurcations in such
a way that two or more eigenvalues and the corresponding wave
functions coalesce at critical values of external parameters.  E.g.\
in condensates without long-range interactions a stable and an
unstable state are created in a tangent bifurcation when the
scattering length of the contact interaction is varied.  At the
critical point the coalescing states show the properties of an
exceptional point.  In dipolar condensates fingerprints of a pitchfork
bifurcation have been discovered by Rau \etal [Phys.\ Rev.\ A,
81:031605(R), 2010].  We present a method to uncover all states
participating in a pitchfork bifurcation, and investigate in detail
the signatures of exceptional points related to bifurcations in
dipolar condensates.  For the perturbation by two parameters, viz.\
the scattering length and a parameter breaking the cylindrical
symmetry of the harmonic trap, two cases leading to different
characteristic eigenvalue and eigenvector patterns under cyclic
variations of the parameters need to be distinguished.  The observed
structures resemble those of three coalescing eigenfunctions obtained
by Demange and Graefe [J.\ Phys.\ A, 45:025303, 2012] using
perturbation theory for non-Hermitian operators in a linear model.
Furthermore, the splitting of the exceptional point under symmetry 
breaking in either two or three branching singularities is examined.
Characteristic features are observed when one, two, or three
exceptional points are encircled simultaneously.
\end{abstract}

\pacs{03.75.Kk, 03.65.Vf}
%\maketitle

% Abbreviations
\acrodef{BEC}{Bose-Einstein condensate}
\acrodef{GPE}{Gross-Pitaevskii equation}
\acrodef{TDVP}{time-dependent variational principle}

\section{Introduction}
\label{sec:introduction}
The macroscopic occupation of the bosonic ground state of ultracold
quantum gases has been predicted by Bose and Einstein, and at least
since the first realisation of Bose-Einstein condensates (BECs)
\cite{And95a,Bra95a,Dav95a} they are a central part of experimental
and theoretical atomic physics.
The condensates are typically held in a harmonic trap.
The particles in the trap interact via short-range contact
interactions determined by the s-wave scattering length $a$, which can
be varied experimentally with the help of Feshbach resonances.
In addition long-range interactions exist, e.g., in dipolar
condensates \cite{Lahaye09} or in condensates with an attractive $1/r$
interaction \cite{ODe00,Pap07}.

In a mean-field approach the effective one-particle wave function of
the condensate is described by the nonlinear Gross-Pitaevskii equation
(GPE).
The nonlinearity of the GPE allows for a variety of phenomena
which are impossible in linear quantum systems with Hermitian operators.
For example, the extended Gross-Pitaevskii equation for a BEC without
long-range interactions or with an attractive $1/r$ interaction has a
second solution which emerges together with the ground state in a
tangent bifurcation \cite{Pap07}.
At the bifurcation point both states coalesce, i.e., the energies and
the wave functions are identical.
It has been shown \cite{Car08a} that the bifurcation point has the
properties of an ``exceptional point''
\cite{Kato66,Hei90,Hei99,Heiss08,Hei12}.
Such points can appear in systems described by non-Hermitian matrices
which depend on a two-dimensional parameter space.
The exceptional points are critical points in the parameter space
where both the eigenvalues and the eigenvectors of the two states pass
through a branch point singularity and become identical.
There is only one linearly independent eigenvector of the two states
at an exceptional point.

Bifurcation scenarios become even more substantial in dipolar
condensates.
With a simple variational approach a tangent bifurcation between the
ground state and an unstable excited state was observed in
\cite{Koeberle09a}, and it was shown that the bifurcation point has
the properties of an exceptional point.
However, using an extended variational approach with coupled Gaussians
for the condensate wave function the ground state is created unstable
and only becomes stable at an increased value of the scattering length
\cite{Rau10c}.
Although no bifurcating states have been directly computed in
\cite{Rau10c}, the stability change indicates the existence of a
pitchfork bifurcation, which means that three states coalesce at the
bifurcation point.

The detailed investigation of pitchfork bifurcations and signatures of
three coalescing eigenfunctions in dipolar condensates is the
objective of this paper.
We present a method which exploits the symmetry breaking of the
external harmonic trap and allows us to reveal all three states
involved in the pitchfork bifurcation.
The energies and the eigenfunctions of the three states coalesce at
the bifurcation point, however, the degeneracies are lifted when
the scattering length is varied or the axial symmetry of the harmonic
trap is broken.

The signatures of three coalescing eigenfunctions have been
investigated by Demange and Graefe \cite{Dem12} using perturbation
theory for a model with linear but non-Hermitian operators.
It was shown that two types of parameter perturbations need to be
distinguished, which are related to a cubic root and a square root
branching singularity at the exceptional point and lead to
characteristic patterns under cyclic variation of the parameters.
With an analytic continuation of the {\em nonlinear} GPE in the two
parameters, viz.\ the scattering length and a parameter breaking the
axial symmetry of the harmonic trap, we show that the signatures of
the three states of a dipolar BEC which coalesce in the pitchfork
bifurcation exactly resemble those obtained in the {\em linear} model
\cite{Dem12} with complex non-Hermitian matrices.
Both, the cubic root and the square root branching singularity are
observed when encircling the exceptional point either in the symmetry
breaking parameter or in the scattering length.
Furthermore, we will discuss the behaviour of the eigenvalues and
eigenvectors when both control parameters are simultaneously varied.

\section{Variational approach to dipolar condensates}
\label{sec:tdvp}
An extended variational ansatz with coupled Gaussian functions has
been used by Rau \etal \cite{Rau10a,Rau10b} to compute numerically
accurate solutions of the GPE for condensates with dipolar
interactions.
Contrary to the numerical imaginary time evolution of states on grids
the variational approach allows for the calculation of not only the
ground state but also excited states, which is a prerequisite for the
investigation of bifurcation scenarios and exceptional points.
In this section we briefly recapitulate the variational method and
introduce the analytic continuation of the GPE, which then allows for
the encircling of exceptional points in the complex parameter space.

We use the time-dependent GPE in dimensionless and particle number
scaled units (see \cite{Koeberle09a}),
\begin{equation}
 H \psi(\vec r, t) = [ -\Delta + V ] \psi(\vec r, t) =
  \ii \frac{\partial}{\partial t} \psi(\vec r, t) \; ,
\label{eq:GPE}
\end{equation}
to describe dipolar BECs in the mean-field approximation.
The total potential
\begin{equation}
 V = V_{\rm t} + V_{\rm c} + V_{\rm d}
\label{eq:V}
\end{equation}
is composed of three parts.
The external trap is described by the harmonic potential
\begin{equation}
 V_{\rm t} = \gamma_x^2 x^2 + \gamma_y^2 y^2 + \gamma_z^2 z^2 \; ,
\end{equation}
where the $\gamma_{x,y,z}$ determine the strength of the trap in the
$x$, $y$, and $z$ directions.
The dipolar condensates are typically held in an axisymmetric trap
with $\gamma_x=\gamma_y$.
In order to describe the trap frequencies in a way that represents the
trap geometry and to allow us to change the trap symmetry by modifying
only one parameter we will use the parametrisation
\begin{equation}
 \gamma_x = \bar\gamma (1+s)^{1/2} \lambda^{-1/3} \; , \quad
 \gamma_y = \bar\gamma (1+s)^{-1/2} \lambda^{-1/3} \; , \quad
 \gamma_z = \bar\gamma \lambda^{2/3}
\label{eq:Vt}
\end{equation}
where $\bar\gamma$ is the average trap frequency and $\lambda$ the
ratio between the trap frequency in $z$-direction and the trap
frequencies in the $x,y$-plane.
The asymmetry parameter $s$ breaks, for $s\ne 0$, the axial symmetry
of the trap, and is one of the two parameters used below to examine
the signatures of exceptional points.
The potential
\begin{equation}
 V_{\rm c} =  8 \pi a | \psi(\vec r, t) |^2
\label{eq:Vc}
\end{equation}
describes the short-range contact interactions resulting from s-wave
scattering between particles with $a$ the (scaled) scattering length.
The scattering length $a$ is the second parameter used to examine the
signatures of exceptional points.
Finally, the potential
\begin{equation}
 V_{\rm d} = \int \dd^3 r' 
 \frac{1-3\cos^2\vartheta}{|\vec r - \vec{r'}|^3} |\psi(\vec r', t)|^2
\label{eq:Vd}
\end{equation}
describes the dipole-dipole interaction between particles with a
magnetic moment, with $\vartheta$ the angle between the vector
$\vec r - \vec r'$ and the $z$-axis along which all dipoles are
aligned by an external magnetic field.

In order to solve the GPE (\ref{eq:GPE}) we use the time-dependent
variational principle (TDVP) \cite{McLachlan}.
The idea of the TDVP is to use an ansatz for the wave function
depending on the set of variational parameters $\vec z$ and then
choose these parameters in such a way that the quantity
$I=\|H\psi-\ii\dot\psi\|^2$ is minimised.
We use a superposition of $N$ Gaussian functions
\begin{equation}
 \psi(\vec r, \vec z) = \sum_{k=1}^N g^k = \sum_{k=1}^N \ee^{-
 \left( A_x^k x^2 + A_y^k y^2 + A_z^k z^2 + \gamma^k \right) }
\label{eq:ansatz}
\end{equation}
as an ansatz for the wave function, where the $A_\sigma^k$ are the
complex width parameters of the $k$th Gaussian in $\sigma$ direction,
and the real and imaginary parts of the $\gamma^k$ describe the phase
and the amplitude of the $k$th Gaussian, respectively.
The variational parameters are combined in the complex vector
\begin{equation}
 \vec z = \left(\gamma^1, \dots, \gamma^N, A_x^1, \dots, A_x^N, A_y^1,
 \dots, A_y^N,  A_z^1, \dots, A_z^N \right) \in \mathbb{C}^{4N} \; .
\label{eq:parameter}
\end{equation}
The application of the TDVP (for details of the derivations see
\cite{Rau10a}) leads to the following equations of motion.
The time derivatives of the variational parameters are given by
$\dot{\vec z}=\vec h(\vec z)$, or in components
\begin{eqnarray}
 \dot \gamma^k &= 2 \ii \left( A_x^k + A_y^k + A_z^k \right)
  + \ii v_0^k \; , \nonumber \\
 \dot A_\sigma^k &= -4\ii (A_\sigma^k)^2 + \ii v_\sigma^k \quad \text{with} \quad
 \sigma=x,y,z \; .
\label{eq:timederivative}
\end{eqnarray}
The quantities $v_0^k$ and $v_\sigma^k$ in equation
(\ref{eq:timederivative}) can be written as a vector
$\vec v=(v_0^1,\dots,v_0^N,v_x^1,\dots,v_x^N,v_y^1,\dots,v_y^N, 
v_z^1,\dots,v_z^N)$, which is obtained by solving the linear system of
equations
\begin{equation}
  M \vec v = \vec r \; .
\label{eq:linsystem}
\end{equation}
The matrix $M$ has the form
\begin{equation}
 M = \left(\begin{array}{cccc}
  \left(1\right)_{lk} &
  \left(x^2\right)_{lk} &
  \left(y^2\right)_{lk} &
  \left(z^2\right)_{lk} \\
  \left(x^2\right)_{kl} &
  \left(x^4\right)_{lk} &
  \left(x^2y^2\right)_{lk} &
  \left(x^2z^2\right)_{lk} \\
  \left(y^2\right)_{kl} &
  \left(x^2y^2\right)_{kl} &
  \left(y^4\right)_{lk} &
  \left(y^2z^2\right)_{lk} \\
  \left(z^2\right)_{kl} &
  \left(x^2z^2\right)_{kl} &
  \left(y^2z^2\right)_{kl} &
  \left(z^4\right)_{lk}
           \end{array}\right)
\label{eq:matrix_M}
\end{equation}          
with the submatrices
\begin{equation}
 (O)_{lk}=\langle g^l|O|g^k\rangle
\end{equation}
for $k,l=1,\dots,N$ and the operators
$O=1,x^2,y^2,z^2,x^2y^2,x^2z^2,y^2z^2,x^4,y^4,z^4$.
The components of the right-hand side vector
$\vec r=(r_1^1,\dots,r_1^N,r_x^1,\dots,r_x^N,r_y^1,\dots,r_y^N,
r_z^1,\dots,r_z^N)$ in equation (\ref{eq:linsystem}) read
\begin{equation}
 r_f^l = \sum_{k=1}^N \langle g^l|f^2\, V|g^k\rangle
\label{eq:r_components}
\end{equation}
for $f=1,x,y,z$ and $l=1,\dots,N$.
All components of the matrix $M$ and the integrals for the vector
$\vec r$ in equation (\ref{eq:r_components}) are listed in
\ref{sec:app}.
The components of the matrix $M$ can be expressed analytically.
The vector $\vec r$ is more complicated due to the dipole interaction.
The components
\begin{eqnarray}
 \langle g^k | f^2\,V_{\rm d} | g^l \rangle
            ~~\text{with}~ f = 1, x, y, z
\label{eq:element_difficult}
\end{eqnarray}
lead to expressions which contain elliptic integrals of the form
\begin{equation}
 R_D(x, y, z) = \frac{3}{2} \int_0^\infty 
 \frac{\dd t}{\sqrt{(x+t)(y+t)(z+t)^3}} \; .
\label{eq:elliptic_integral}
\end{equation}
These integrals can be calculated numerically using the Carlson
algorithm \cite{Carlson}.

The stationary states are obtained as fixed points of the equations of
motion (\ref{eq:timederivative}), i.e.\ $\dot\gamma^k-\ii\mu=0$ and
$\dot A_\sigma^k = 0$ with $\mu$ the chemical potential,
$\sigma=x,y,z$, and $k=1,\dots,N$.
The set of the variational parameters $\gamma$ can be reduced to
$\tilde\gamma^k=\gamma^k-\gamma^1$ for $k=2,\dots,N$ by using the facts
that the wave function is normalised to $\langle\psi|\psi\rangle=1$
and a global phase of the wave function is arbitrary.
The conditions for a stationary state then simplify to
\begin{eqnarray}
 \dot{\tilde\gamma}^k & = & \dot\gamma^k - \dot\gamma^1 = 0 \; ;
  \quad k=2,\dots,N \; , \nonumber \\
 \dot A_x^k & = & \dot A_y^k = \dot A_z^k = 0 \; ; \quad k=1,\dots,N \; .
\label{eq:rootsearch}
\end{eqnarray}
The root search for the fixed points is performed using the
Newton-Raphson algorithm.

The mean-field energy of a stationary state is given by
\begin{equation}
  E_{\rm mf} = \langle \psi | -\Delta + V_{\rm t} +
      \textstyle{\frac{1}{2}}(V_{\rm c} + V_{\rm d}) | \psi \rangle \; ,
\label{eq:emf}
\end{equation}
and can be used to examine some properties of the state.
However, the typical bifurcation behaviour is revealed even more
clearly by, e.g.\ the expectation values $w$ of the operator
$x^2-y^2$, viz.
\begin{equation}
  w = \langle \psi | x^2 - y^2 | \psi \rangle \; ,
\label{eq:w}
\end{equation}
and therefore this value is used below in most cases to reveal the
bifurcations and to analyse the signatures of the exceptional points.
For the presentation of the splittings we use the distance from their
mean values,
\begin{eqnarray}
\label{eq:traceless_e}
 \Delta E_{{\rm mf},j} &=& E_{{\rm mf},j} - \frac{1}{N_{\rm s}}
 \sum_{i=1}^{N_{\rm s}} E_{{\rm mf},i} \; , \\
\label{eq:traceless_w}
 \Delta w_j &=& w_j - \frac{1}{N_{\rm s}} \sum_{i=1}^{N_{\rm s}} w_i \; ,
\end{eqnarray}
i.e.\ the mean values of the quantities of all $N_{\rm s}$ states
participating in the bifurcation (with $N_{\rm s}=2$ for a tangent
bifurcation and $N_{\rm s}=3$ for a pitchfork bifurcation) are subtracted.
The index $j$ indicates the $j$th state $|\psi_j\rangle$ of the
bifurcation.
With the new values the level-splittings at the bifurcations (which are
small compared to the absolute values) can be clearly observed.

The linear stability of stationary states is analysed by calculating
the eigenvalues of the Jacobi matrix
\begin{equation}
 J_{ij} = \frac{\partial \dot{\tilde z}_i}{\partial \tilde z_j} \; ,
\label{eq:Jacobi}
\end{equation}
which is obtained with the variational parameters split into their
real and imaginary parts, viz.
\begin{equation}
 \tilde{\vec z} = (\Re z_1, \Im z_1, \dots, \Re z_{4N}, \Im z_{4N})
 \in \mathbb{R}^{8N} \; .
\label{eq:ztilde}
\end{equation}
If all eigenvalues are purely imaginary, the state is stable,
otherwise it is unstable.
The eigenvalues occur in pairs $\pm\lambda$.
If one pair of eigenvalues has a nonzero real part the state is
unstable in the directions of the associated eigenvectors in the
parameter space.
If two pairs of eigenvalues have a nonzero real component the state is
unstable in four directions, and so on.

\subsection{Analytic continuation of the GPE}
\label{sec:analytic_cont}
The parameter $s$ in equation (\ref{eq:Vt}), which breaks the axial
symmetry of the harmonic trap, and the scattering length $a$ in
equation (\ref{eq:Vc}) are the two, physically real, parameters used
to control the system.
To investigate a bifurcation point in equation \eqref{eq:timederivative}
for the signatures of an exceptional point it is necessary to encircle
the critical value in the complex plane.
Therefore, the GPE (\ref{eq:GPE}) must be continued analytically with
respect to the parameters $s$ and $a$ in such a way that observables,
such as the mean-field energy $E_{\rm mf}$ or the value of $w$, can
become complex.
As the variational parameters $\vec z\in\mathbb{C}^{4N}$ are complex
parameters it is necessary to split the components into their real and
imaginary parts, and to start with the real parameters
$\tilde{\vec z}\in\mathbb{R}^{8N}$ introduced in equation (\ref{eq:ztilde}).
The real components of the vector $\tilde{\vec z}$ can now be
continued analytically as
\begin{equation}
 \tilde z_l = \tilde z_l^R + \kk \tilde z_l^I
\label{eq:ztilde_compl_cont}
\end{equation}
with the real parameters $z_l^R$ and $\tilde z_l^I$ for
$l=1,\dots,8N$, and the imaginary unit $\kk$ defined by $\kk^2=-1$.
It is important to distinguish $\kk$ from the imaginary unit $\ii$ for
the following reason.
The complex conjugate of the wave function $\psi(\vec r,\vec z)$ in
equation (\ref{eq:ansatz}) is obtained by the replacement $\ii\to -\ii$
and thus $\vec z\to\vec z^\ast$, however, the sign of $\kk$ in the
complex continued vector $\tilde z$ in equation (\ref{eq:ztilde_compl_cont})
must {\em not} be changed.
Formally, the variational parameters $\vec z$ can be written as
bicomplex numbers
\begin{eqnarray}
 z_l &= z_l^R + \ii\, z_l^I
     = (z_{l,1}+\kk\, z_{l,2}) + \ii\, (z_{l,3}+\kk\, z_{l,4}) \nonumber \\
     &= z_{l,1} + \kk\, z_{l,2} + \ii\, z_{l,3} - \jj\, z_{l,4}
\label{eq:zbicompl}
\end{eqnarray}
with $\ii^2 = \kk^2 = -1$, $\jj^2 = 1$, $\ii\jj = \jj\ii = \kk$,
$\jj\kk = \kk\jj = \ii$, and $\ii\kk = \kk\ii = -\jj$.
The complex conjugate wave function in this description has the form
\begin{eqnarray}
 \psi^\ast(\vec r) = \sum_{l=1}^N \ee^{-\left( A_x^{l\ast} x^2 +
     A_y^{l\ast} y^2 + A_z^{l\ast} z^2 + \gamma^{l\ast} \right) } \; ,
\end{eqnarray}
with $A_\sigma^{l\ast}=(A_{\sigma,1}^l + \kk A_{\sigma,2}^l)
-\ii(A_{\sigma,3}^l + \kk A_{\sigma,4}^l)$ and
$\gamma^{l\ast}=(\gamma_1^l+\kk \gamma_2^l)-\ii(\gamma_3^l+\kk\gamma_4^l)$.
For complex parameters $s$ and $a$ the analytically continued
equations of motion (\ref{eq:timederivative}) can now be set up for
the variational parameters extended to bicomplex numbers.
The advantage of using bicomplex numbers is that the rules for
calculations can be easily implemented in computer algorithms, e.g.\
by introducing a new data type and overloading operators.
In particular, Carlson's algorithm \cite{Carlson} for the computation
of the elliptic integrals $R_D(x,y,z)$ in equation
(\ref{eq:elliptic_integral}) can be used with nearly no modifications
for real, complex, and even bicomplex numbers $x$, $y$, and $z$.
The linear set of equations (\ref{eq:linsystem}) can be solved either
directly with bicomplex numbers or the components of the bicomplex
matrix $M$, and the vectors $\vec v$ and $\vec r$ can be split into two
complex numbers to set up an ordinary complex linear system of
equations of twice the dimension.

The stationary states are obtained as roots of equations
(\ref{eq:rootsearch}) extended to bicomplex variational parameters.
Note that with the analytic continuation observables such as the
mean-field energy $E_{\rm mf}$ or the value of $w$ are not necessarily
real but can become complex valued, e.g.\ $w=w_1+\kk w_2$ with the
$\ii$ and $\jj$ components $w_3=w_4=0$.
In the following we use the notation $\Re w=w_1$ and $\Im w=w_2$ in
such cases.

\section{Bifurcations of the stationary states}
\label{sec:Bifurcations}
We study the bifurcation scenarios of the stationary states of a
dipolar BEC in a harmonic trap for the case that the average trap
frequency $\bar\gamma$, the ratio $\lambda$, and the symmetry breaking
parameter $s$ in equation (\ref{eq:Vt}) are kept constant and the
scattering length $a$ is varied.
\begin{figure}
\includegraphics[width=0.99\textwidth]{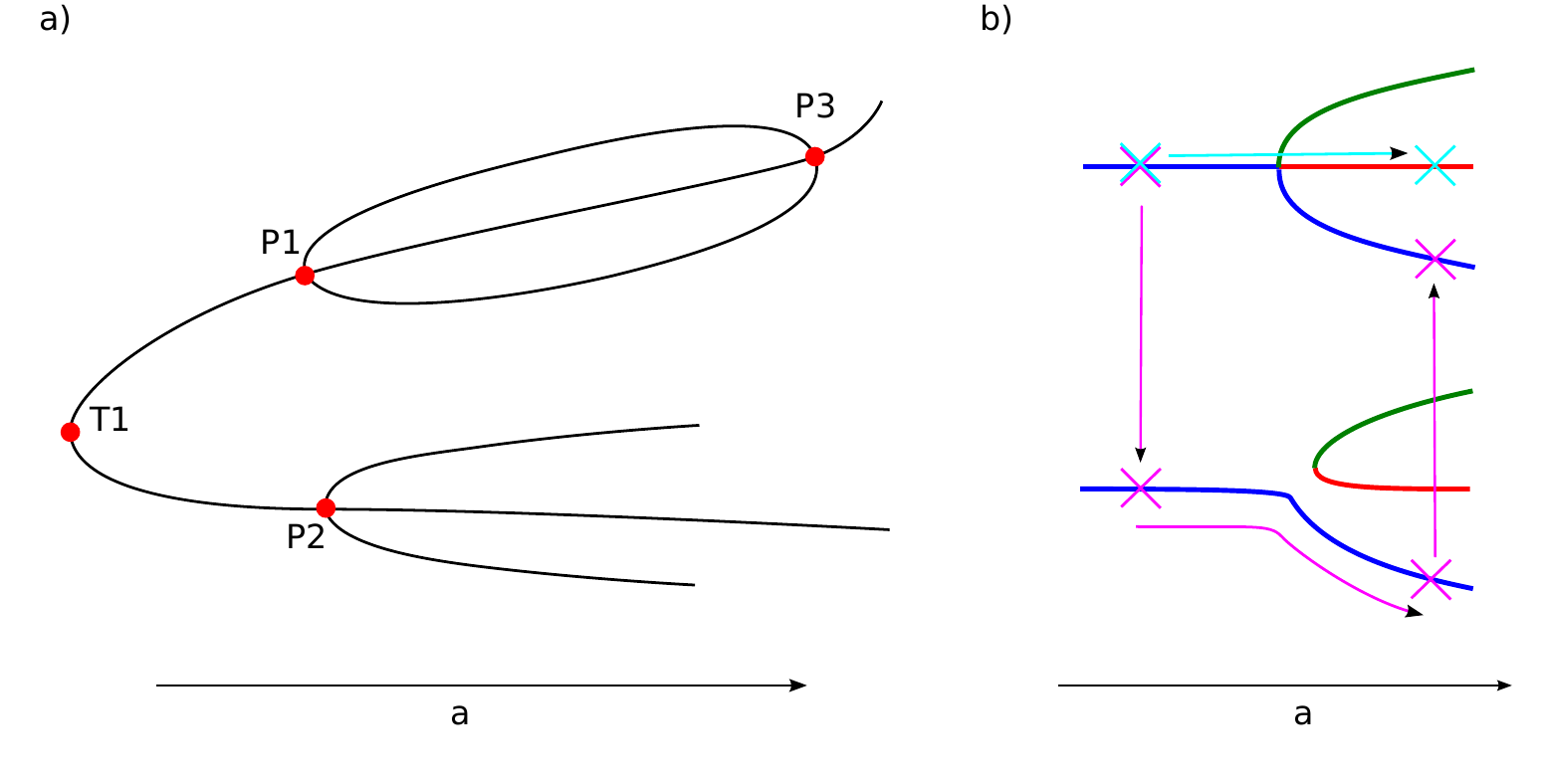}
\caption{(a) Sketch of the bifurcation scenario of a dipolar BEC in an
  axisymmetric trap.  Two states emerge in a tangent bifurcation at T1.
  The two branches undergo pitchfork bifurcations at P1 and P2.  The
  three states created in the pitchfork bifurcation P1 merge again in
  an inverse pitchfork bifurcation at P3.
  (b) Scheme how to find the branching states of the pitchfork
  bifurcation: Starting at $s = 0$ on the central state marked by a
  cross in the left-hand upper part of the figure the axial symmetry
  of the trap is broken (cross in the lower part of the figure), then
  the state is followed adiabatically by increasing the scattering
  length, and finally the axial symmetry of the trap is restored to
  end up with a state on one of the symmetry breaking branches in the
  upper part of the figure.}
\label{fig1}
\end{figure}
A sketch of the bifurcation scenario is given in figure~\ref{fig1}(a).
The ground state and an excited state emerge in a tangent bifurcation
(marked T1 in figure~\ref{fig1}(a)) at a critical scattering length
$a=a_{\rm T1}$ \cite{Koeberle09a,Rau10b}.
When the scattering length is increased both the ground and the
excited state can undergo further bifurcations indicated P1, P2, and
P3 in figure~\ref{fig1}(a).

\subsection{Bifurcations of the excited state}
\label{sec:bif_excited}
We start with the investigation of the stationary states using a
single Gaussian function as a variational ansatz for the wave
function.
The trap parameters are set to $\bar\gamma_1=34000$, $\lambda_1=6$, and
$s=0$ for a harmonic potential with axial symmetry, which is in the
region where dipolar condensates with chromium atoms have been
realised experimentally \cite{Koch08,Lahaye09}.
The bifurcation scenario drafted in figure~\ref{fig1}(a) (without P2)
emerges.
At the scattering length $a_{\rm T1}=-0.019$ a stable ground state and
an unstable excited state are created in a tangent bifurcation (T1).
Both wave functions are axisymmetric around the $z$ axis.
At the scattering length $a_{\rm P1}=-0.0079$ the excited state
undergoes a pitchfork bifurcation P1, where its stability properties
change from unstable for one pair of stability eigenvalues to unstable
for two pairs of stability eigenvalues.
If the scattering length is further increased all three states which
emerge in the pitchfork bifurcation are recombined in an inverse
pitchfork bifurcation at $a_{\rm P3}=0.129$.
At the bifurcation points not only the mean-field energies and the
expectation values of the operator $x^2-y^2$ but also the wave
functions coalesce.

\begin{figure}
\includegraphics[width=\textwidth]{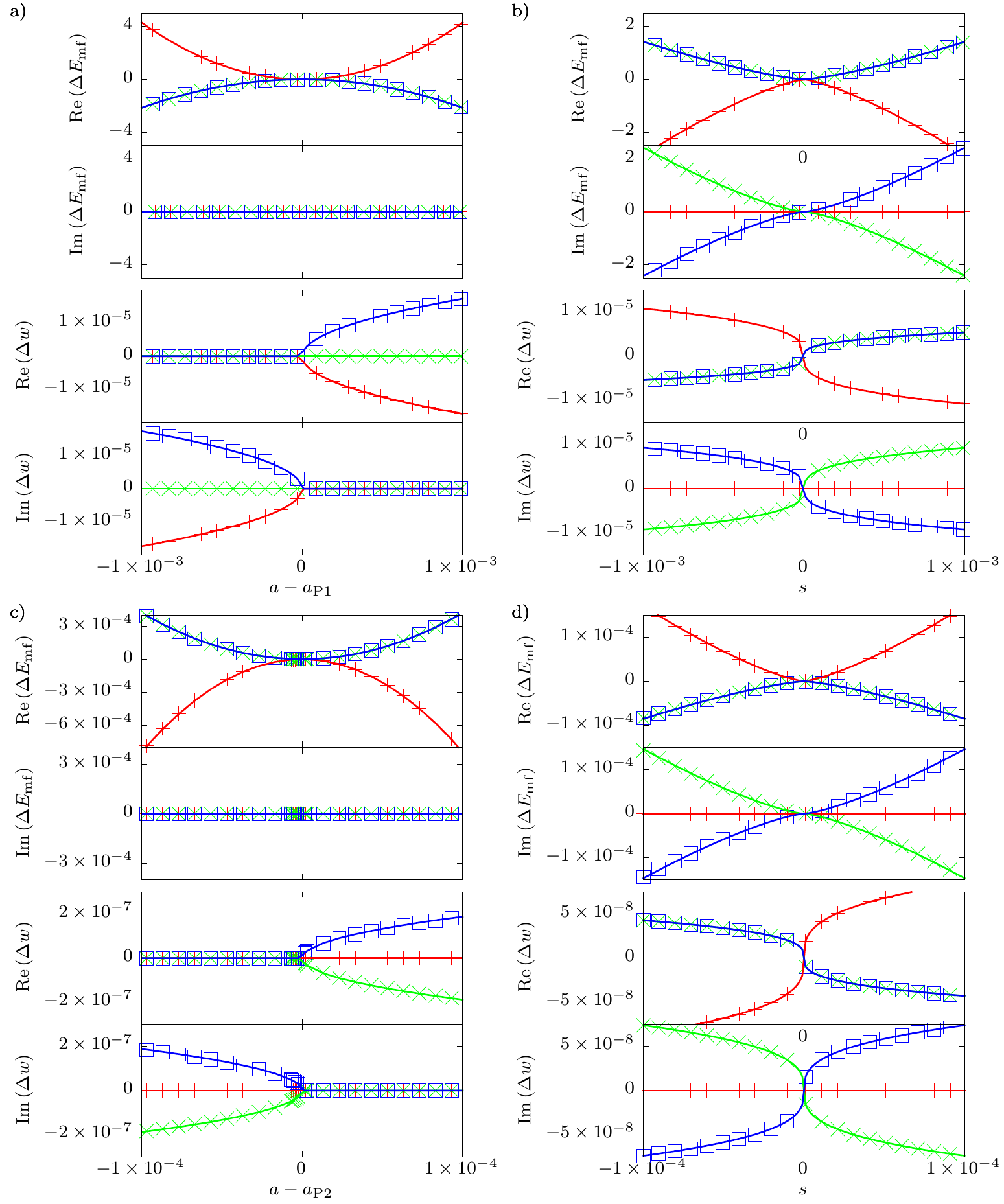}
\caption{(a)-(b) Level splittings of the mean-field energy $E_{\rm mf}$
  and the values $w=\langle\psi|x^2-y^2|\psi\rangle$ around the
  pitchfork bifurcation P1 of the excited state. The symbols and solid
  lines mark the results of the GPE using a variational ansatz with a
  single Gaussian function and the linear model introduced in section
  \ref{sec:matrix_model}, respectively. In (a) the scattering length
  is varied for a symmetric trap ($s=0$) and in (b) the asymmetry
  parameter $s$ is varied while the scattering length is kept constant
  ($a=a_{\rm P1}$). (c)-(d) Results for the pitchfork bifurcation P2
  of the ground state. Here, the GPE was solved using an ansatz with
  six coupled Gaussians.}
\label{fig2}
\end{figure}

In figure~\ref{fig2}(a)-(b) the detailed results for the pitchfork
bifurcation P1 around the critical scattering length $a_{\rm P1}=-0.0079$
are presented.
Shown are both the real branches and the complex branches obtained by
analytic continuation of the GPE.
The  characteristic shape of a pitchfork bifurcation is evident in the
real part of $\Delta w$ in figure~\ref{fig2}(a), where two new real
branches emerge at the bifurcation point $a_{\rm P1}=-0.0079$.
The new emerging states are unstable with one pair of stability
eigenvalues with nonzero real part, and the corresponding wave
functions break the axial symmetry of the trap.
The two states only differ by a $90^\circ$ rotation around the $z$
axis.
With the analytic continuation two additional complex branches are
found below the bifurcation point as can be seen in the imaginary part
of $\Delta w$ in figure~\ref{fig2}(a).

The existence of a new branch at $a>a_{\rm P1}$ is also visible in
figure~\ref{fig2}(a) for the splitting $\Delta E_{\rm mf}$ of the
mean-field energy.
As the two new states only differ by a $90^\circ$ rotation around the
$z$ axis in the axisymmetric trap the mean-field energy of the new
branch is twofold degenerate.
Furthermore, the splitting $\Delta E_{\rm mf}$ increases quadratically
with $a-a_{\rm P1}$ in contrast to the typical square root behaviour
of a pitchfork bifurcation, and the mean-field energy is real also for
those states obtained with the analytic continuation of the GPE in the
region $a<a_{\rm P1}$, i.e.\ ${\rm Im}\,(\Delta E_{\rm mf})=0$ for all
states in figure~\ref{fig2}(a).

When the scattering length is kept constant at $a=a_{\rm P1}$ and the
asymmetry  parameter $s$ is varied the splittings $\Delta E_{\rm mf}$
and $\Delta w$ show a quite different cubic-root-like behaviour as
illustrated in figure~\ref{fig2}(b).
The splitting is $\Delta E_{\rm mf}\sim s^{4/3}$ for the mean-field
energy and $\Delta w\sim s^{1/3}$ for the expectation values of the
operator $x^2-y^2$, which implies that the splittings are real-valued
only for the central state marked by red plus symbols in
figure~\ref{fig2}(b).
A more detailed discussion of the splittings based on a linear model
with non-Hermitian matrices will be presented in
section~\ref{sec:matrix_model}.

\subsection{Bifurcations of the ground state}
\label{sec:bif_ground}
Using the simple variational approach with a single Gaussian function
($N=1$) the stable ground state emerges in a tangent bifurcation
\cite{Koeberle09a}.
However, the situation becomes more complicated when using an improved
and extended ansatz with coupled Gaussians for the condensate wave
function \cite{Rau10c,Rau10a,Rau10b}.
For $N=6$ Gaussian functions and an external trap with parameters
$\bar\gamma_2=6887$, $\lambda_2=7$, and $s=0$ the ground state is
created unstable in the tangent bifurcation and then becomes stable at
a larger value of the scattering length.
The stability change exhibits fingerprints of a pitchfork bifurcation,
however, the numerical search for the new branches emerging in the
bifurcation is a very nontrivial task in the dimensionally increased
parameter space for coupled Gaussians and thus have not been computed
in \cite{Rau10b}.
Here we present a method based on breaking the axial symmetry of the
harmonic trap to obtain the additional states emerging in the
pitchfork bifurcation.

The method consists of three steps as illustrated in
figure~\ref{fig1}(b).
Starting with an axisymmetric state ($s=0$) at a scattering length
below the bifurcation point the asymmetry  parameter $s$ is varied in
a first step to break the axial symmetry of that state ($s\ne 0$).
In a second step the scattering length $a$ is increased to a value
greater than the critical value of the pitchfork bifurcation.
Using small increments of the scattering length we can follow the
state adiabatically as shown in figure~\ref{fig1}(b).
In a third step the parameter $s$ is changed back to $s=0$ for the
axisymmetric trap, however, the state at the end of this path belongs
to one of the new branches emerging in the pitchfork bifurcation.
The other branch can be reached in the same way with the asymmetry
parameter $s$ changed in the opposite direction.

The states of the new branches break the axial symmetry around the $z$
axis and differ by a $90^\circ$ rotation in a similar way as discussed
for the pitchfork bifurcation of the excited state in
section~\ref{sec:bif_excited}.
The resulting branches of the pitchfork bifurcation P2 in the ground
state branch can be seen as functions of the scattering length
$a-a_{\rm P2}$ and the asymmetry parameter $s$ in figure~\ref{fig2}(c)
and (d), respectively.
Up to a change in sign of the splittings $\Delta E_{\rm mf}$ and 
$\Delta w$ the results qualitatively agree with those of the
bifurcation P1 of the excited state in figure~\ref{fig2}(a)-(b).
The eigenvalues and eigenvectors of all three states coincide at the
critical scattering length $a_{\rm P2}=-0.0036$ of the pitchfork
bifurcation.

\subsection{Linear model with non-Hermitian matrices}
\label{sec:matrix_model}
The bifurcations and exceptional points in dipolar condensates result
from the {\em nonlinearity} of the GPE (\ref{eq:GPE}).
Nevertheless, we now introduce a {\em linear} model with non-Hermitian
matrices which can reproduce the pitchfork bifurcations shown in
figure~\ref{fig2} and the level splittings of both the mean-field
energies $\Delta E_{{\rm mf},j}$ and the values $\Delta w_j$ as defined
in equations (\ref{eq:traceless_e}) and (\ref{eq:traceless_w}),
respectively.
Furthermore, the linear matrix model can  describe the observed
structures when an exceptional point related to one of the pitchfork
bifurcations is encircled in either the asymmetry parameter $s$ or the
scattering length $a$ as discussed in section~\ref{sec:EPs}, and is
even valid for the two parameter perturbations discussed in
section~\ref{sec:2-par-pert}.

The values of $\Delta w_j$ are obtained as eigenvalues of the
non-Hermitian matrix
\begin{equation}
  Q_w = c_w \left(
    \begin{array}{ccc}
      0 & 1 & 0 \\  \tilde a & 0 & 1 \\  s & 0 & 0 
    \end{array} \right) \; ,
%  = Q_0 + \tilde a\, P_a + s\, P_s
\label{eq:matrixmodel}
\end{equation}
where $s$ is the asymmetry parameter, and
\begin{equation}
  \tilde a = c_a (a - a_{\rm cr})
\end{equation}
is the rescaled and shifted scattering length so that the critical
value $a_{\rm cr}$ of the bifurcation is at $\tilde a=0$.
Equation (\ref{eq:traceless_w}) ensures that the matrix $Q_w$ is
traceless, and $c_w$ and $c_a$ are real parameters, which depend on
the values $\bar\gamma$ and $\lambda$ of the trap potential and are
determined by a least-squares fit using the data obtained with the
TDVP as described in section~\ref{sec:tdvp}.
For $\tilde a=0$ and $s=0$ the matrix $Q_w$ has the Jordan form 
of an EP3 with eigenvalue $\lambda=0$ (see section~\ref{sec:EPs}).
The solid lines for the real and imaginary part of $\Delta w$ in
figure~\ref{fig2} present the eigenvalues of the matrix $Q_w$ with
$c_a=26.180$ and $c_w=-1.0780\times 10^{-4}$ for the pitchfork
bifurcation P1 and $c_a=104.46$ and $c_w=1.8436\times 10^{-6}$ for the
pitchfork bifurcation P2.
For $s=0$ one eigenvalue is zero and two eigenvalues are
$\sim\sqrt{a-a_{\rm cr}}$, and for $\tilde a=0$ all three eigenvalues
are $\sim s^{1/3}$. 
Obviously, the results of the matrix model perfectly agree with the
exact data marked by symbols in figure~\ref{fig2}.

The matrix model for the mean-field energy is somewhat more complicated.
For the simplicity of the model we require that the matrix elements
are low-order power functions of $\tilde a$ and $s$, and the
eigenvalues are $\sim\tilde a^2$ for $s=0$ and $\sim s^{4/3}$ for
$\tilde a=0$.
Furthermore, we require that for $s\ne 0$ both matrices $Q_w$ and
$Q_E$ always have the same number of degenerate eigenvalues when the
two parameters $\tilde a$ and $s$ are varied
(see section~\ref{sec:2-par-pert} for more details).
The matrix
\begin{equation}
  Q_E = c_E \left(
    \begin{array}{ccc}
      \tilde a^2 & 9 s & 2 \tilde a \\
      16 \tilde a s & \tilde a^2 & 9 s \\
      9 s^2 & 0 & -2 \tilde a^2
    \end{array}\right)
\label{eq:matrixmodel_mf}
\end{equation}
fulfils these requirements.
The real coefficients $c_a$ and $c_E$ are determined by least-squares
fits, and read
$c_a=26.182$ and $c_E=-3136.3$ for the pitchfork bifurcation P1 and
$c_a=101.76$ and $c_E=4.0582$ for the pitchfork bifurcation P2
in figure~\ref{fig2}.
The eigenvalues of the matrix $Q_E$ shown as solid lines in the graphs for
$\Delta E_{\rm mf}$ agree perfectly with the exact data marked by symbols.

\section{Exceptional points}
\label{sec:EPs}
The coalescence of two eigenvalues and the corresponding eigenvectors
is a feature known as ``exceptional point'' \cite{Kato66,Hei90,Hei99},
and such an EP2 can appear in systems described by non-Hermitian
matrices, which depend on at least a two-dimensional parameter space.
When the exceptional point is encircled in the two-dimensional
parameter space the eigenvalues show a characteristic feature of a
square root branching singularity, which is that the two eigenvalues
permute after one cycle in the parameter space, and the initial
configuration of the eigenvalues is obtained only after two cycles in
the parameter space.
The degeneracy of more than two eigenvalues and eigenvectors is
possible, in principle, however, an EP$n$ in general requires the
adjustment of $(n^2+n-2)/2$ parameters \cite{Heiss08}, which implies
e.g.\ that $5$ parameters are necessary for an EP3.

The signatures of coalescing eigenfunctions have been studied by
Demange and Graefe \cite{Dem12} for complex non-Hermitian matrices.
If these matrices are transformed to their Jordan normal form, the
type of the exceptional point is given by the size of the Jordan block.
For example, a block of size three
\begin{equation}
  J_{\rm EP3} = \left(
    \begin{array}{ccc}
      \lambda & 1 & 0 \cr
      0 & \lambda & 1 \cr
      0 & 0 & \lambda
    \end{array} \right)
\label{eq:JordanBlockEP3}
\end{equation}
characterises an EP3.
Now a perturbation can be added to $J_{\rm EP3}$, viz.
\begin{equation}
  J_{\rm EP3}^\ast = J_{\rm EP3} + \Omega P
  ~\text{with}~ P \in \mathbb{C}^{3\times 3} \; .
\end{equation}
If $\Omega$ describes a complex path which encircles the exceptional
point, e.g.\ $\Omega(\varphi)=\ee^{\ii\varphi}$, the behaviour of the
eigenvalues depends on properties of the perturbation.
If the perturbation matrix $P$ is such that the element
\begin{equation}
  P_{31} \ne 0 \, ,
\label{eq:constraint}
\end{equation}
then the typical cubic-root branching singularity expected for an EP3
\cite{Heiss08} is observed, i.e.\ all three eigenvalues and
eigenstates permute.
However, if the condition (\ref{eq:constraint}) is not fulfilled one
might observe the permutation between two states, i.e.\ the signature
of a square-root branching singularity \cite{Dem12}.

In a dipolar BEC with an axisymmetric trap the eigenvalues and the
eigenvectors of three states coincide in a pitchfork bifurcation by
varying only a single parameter, viz.\ the scattering length $a$ of
the contact interaction.
The coalescence of three states by varying only one parameter is
certainly related to an underlying high symmetry of the system.
In this section we investigate in detail the signatures of the
exceptional points occurring in dipolar condensates.

\subsection{Encircling the exceptional points}
We examine the permutation behaviour of the stationary states when the
bifurcation points are enclosed by a parameter path.
Using the analytic continuation of the GPE introduced in
section~\ref{sec:analytic_cont} or the matrix model of
section~\ref{sec:matrix_model} the exceptional point can be encircled
either in the complex continued scattering length $a$,
\begin{equation}
  a(\varphi) = a_{\rm cr} + r_a \ee^{\kk \varphi}
\label{eq:circle_a}
\end{equation}
or in the complex continued asymmetry parameter $s$,
\begin{equation}
  s(\varphi) = r_s \ee^{\kk \varphi} \; ,
\label{eq:circle_s}
\end{equation}
with the imaginary unit $\kk$ introduced in
section~\ref{sec:analytic_cont}.

The tangent bifurcation marked T1 in figure~\ref{fig1}(a) has already
been studied \cite{Koeberle09a}.
If the complex scattering length follows the path in
equation~(\ref{eq:circle_a}) around the critical scattering length
$a_{\rm T1}=-0.019$, the two states participating in the bifurcation
permute in agreement with previous results \cite{Koeberle09a}.
This behaviour is typical for an EP2.

\begin{figure}
\includegraphics[width=\textwidth]{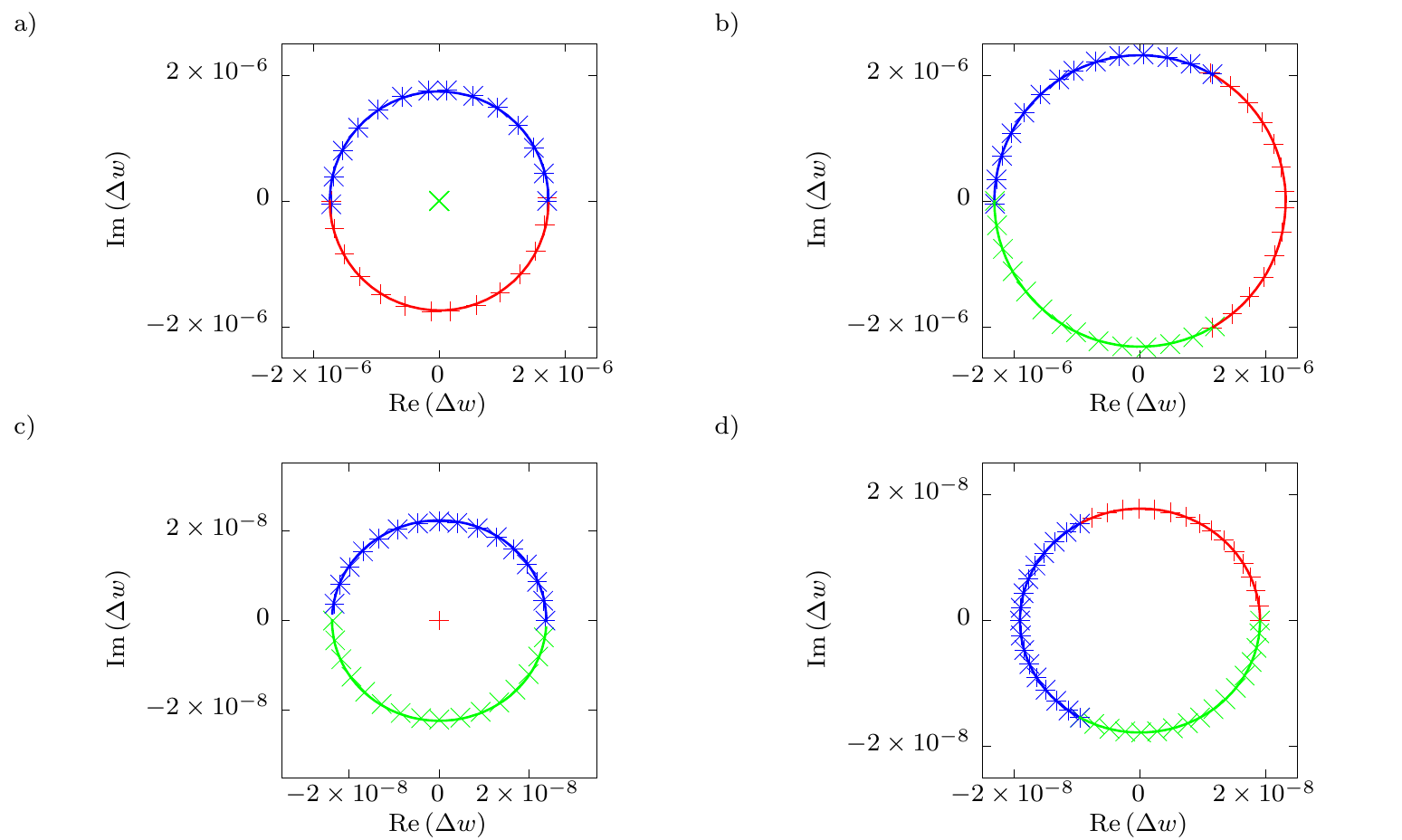}
\caption{Encircling of the pitchfork bifurcation P1 of the excited
  state (a) in the complex scattering length $a$ with the radius
  $r_a=10^{-5}$ and (b) in the complex asymmetry parameter $s$ with the
  radius $r_s = 10^{-5}$. The permutation of only two states in (a)
  and all three states in (b) is clearly exhibited. (c) and (d): Same
  results for the pitchfork bifurcation P2 of the ground state with
  $r_a = 1.5 \times 10^{-5}$ and $r_s=10^{-6}$, respectively. The
  results of the matrix model (solid lines) agree perfectly with the
  exact data (symbols).}
\label{fig3}
\end{figure}

We now investigate the pitchfork bifurcation of the excited state at
$a_{\rm P1}=-0.0079$.
When the exceptional point is encircled by varying the scattering
length along the path of equation~(\ref{eq:circle_a}) with the radius
$r_a=10^{-5}$ the three states behave as visualised in
figure~\ref{fig3}(a) for the expectation values $w$ of the operator
$x^2-y^2$.
Obviously, the figure does not show the permutation of three states,
which is typical for a cubic-root branching singularity of an EP3.
Rather, only the two states emerging in the bifurcation permute, which
is the scenario discussed in \cite{Dem12} for the special case that
the condition (\ref{eq:constraint}) is not fulfilled.
Nevertheless, the typical permutation of three states is observed when
the asymmetry parameter $s$ is varied along the path of
equation~(\ref{eq:circle_s}) with radius $r_s=10^{-5}$.
By following this path the values of the operator $w$ show the
permutation plotted in figure~\ref{fig3}(b).
All three states permute in the way as expected for an EP3
\cite{Heiss08} and discussed in \cite{Dem12} for the generic case that
the condition (\ref{eq:constraint}) is fulfilled.
The different permutation behaviour for both control parameters
becomes evident in the matrix model introduced in
section~\ref{sec:matrix_model}.
For $s=0$ the matrix element $P_{31}$ of the perturbation in equation
(\ref{eq:matrixmodel}) vanishes and thus a square root behaviour is
expected for two of the eigenvalues \cite{Dem12}, i.e.\ one eigenvalue
is constant, $\Delta w_1=0$, and two eigenvalues follow the paths
$\Delta w_{2,3}\sim\ee^{\kk\varphi/2}$ and permute after one circle
around the exceptional point.
When the exceptional point is encircled in the complex $s$ parameter,
the condition $P_{31}\ne 0$ is fulfilled, and the three eigenvalues
are $\Delta w_j\sim\ee^{\kk\varphi/3}$, resulting in the typical
permutation of all three states as expected for an EP3.
  
The pitchfork bifurcation P2 occurring in the ground state for an
ansatz with $N=6$ coupled Gaussian functions can be analysed in the
same way and exhibits a similar behaviour as the bifurcation of the
excited state.
For the path in the scattering length (equation~(\ref{eq:circle_a}))
with radius $r_a=1.5\times10^{-5}$ encircling the bifurcation at
$a_{\rm P2}=-0.0036$ the values of the operator $w$ show the
permutation of two states (see figure~\ref{fig3}(c)).
However, if the asymmetry parameter $s$ follows the path given in
equation~(\ref{eq:circle_s}) with $r_s=10^{-6}$ all three states
permute as shown in figure~\ref{fig3}(d).

As already discussed above, the mean-field energy near the points P1
and P2 in figure~\ref{fig2} does not show the typical level splittings
expected for a pitchfork bifurcation, and this is also true when the
exceptional points are encircled.
The analysis of the matrix model $Q_E$ for the mean-field energy in
equation (\ref{eq:matrixmodel_mf}) reveals that two eigenvalues
$\Delta E_{\rm mf}$ follow a path $\sim\ee^{2\kk\varphi}$ when the
exceptional point is encircled in the scattering length $a$, i.e.\ one
circle in the scattering length results in two loops in the mean-field
energy.
When the exceptional point is encircled in the asymmetry parameter $s$
all three mean-field energies follow paths
$\sim\ee^{\kk(4/3)\varphi}$, which, however, also means a permutation
of all three states after one circle in $s$.

\subsection{Two parameter perturbations}
\label{sec:2-par-pert}
\begin{figure}
\includegraphics[width=\textwidth]{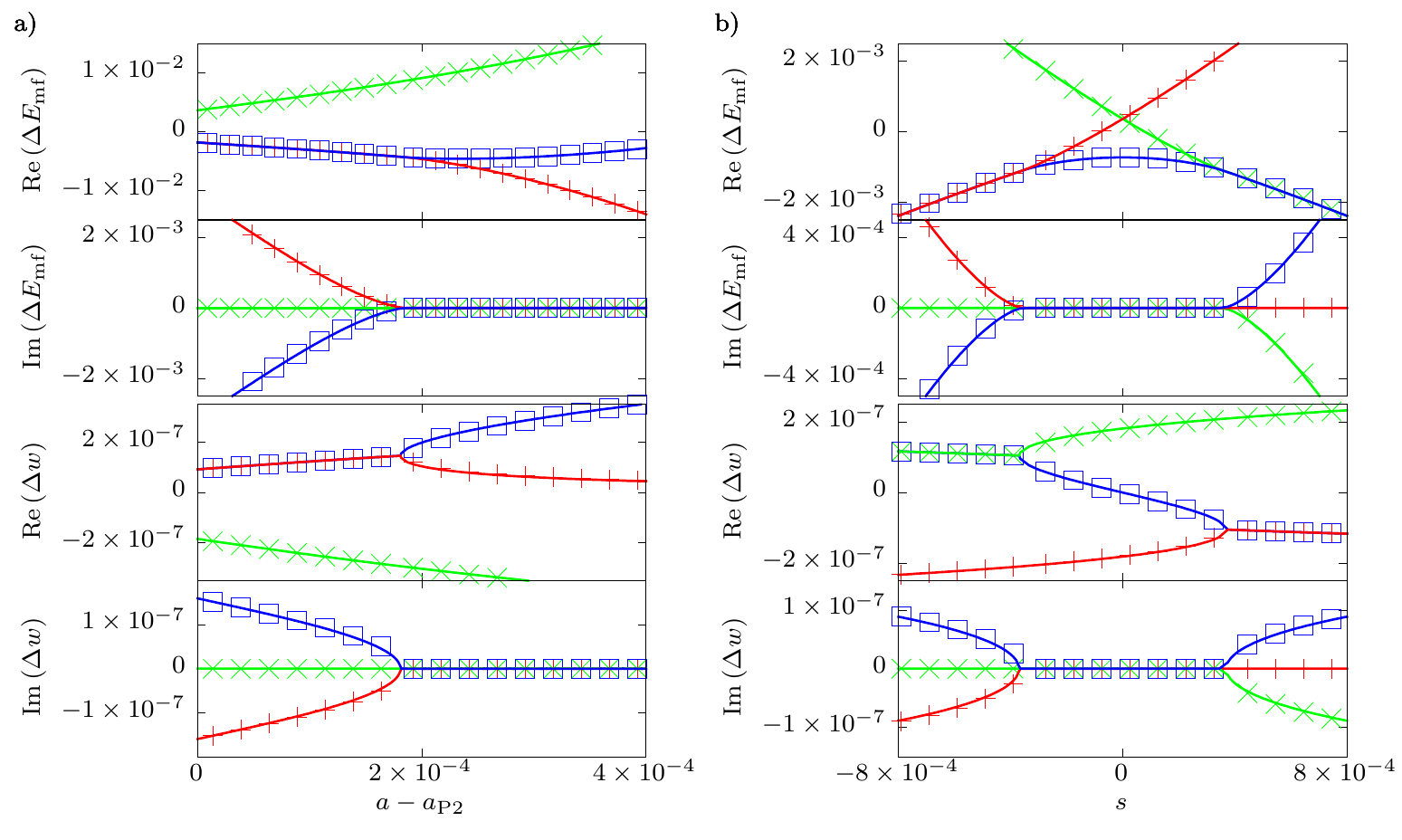}
\caption{Level splittings $\Delta E_{\rm}$ and $\Delta w$ for two
  parameter perturbations of the pitchfork bifurcation P2. In (a) the
  cylindrical symmetry of the trap is broken with $s = -10^{-3}$.
  In (b) the scattering length is chosen such that
  $a-a_{\rm P2} = 9.35\times 10^{-5}$.
  The symbols mark the results obtained using an ansatz of six
  Gaussian functions and agree perfectly with the solid lines obtained
  with the linear matrix model.}
\label{fig4}
\end{figure}

In the previous sections we examined situations where only one of the
parameters $s$ or $a-a_{\rm cr}$ is nonzero.
We now investigate perturbations of the pitchfork bifurcations where
both parameters are nonzero.
Results where either $s$ or $a-a_{\rm cr}$ is set to a constant
nonzero value and the other parameter is varied are presented in
figure~\ref{fig4}.
In figure~\ref{fig4}(a) the symmetry of the trap is broken by setting
$s=-10^{-3}$.
The pitchfork bifurcation does no longer exist and only a tangent
bifurcation between two states remains, as schematically illustrated
in figure~\ref{fig1}(b).
The typical square root behaviour of a tangent bifurcation can be
clearly seen in the real part of $\Delta w$ in figure~\ref{fig4}(a),
the branch below the bifurcation point of the tangent bifurcation
belongs to complex eigenvalues and is obtained only with the analytic
continuation of the GPE.
In figure~\ref{fig4}(b) the scattering length is set to a constant value
$a=a_{\rm P2}+9.35\times 10^{-5}$ and the asymmetry parameter $s$ is varied.
Here, the pitchfork bifurcation is replaced with two tangent
bifurcations located at $s=\pm 3.6\times 10^{-4}$.
Three real states exist in the region between the two tangent
bifurcations, outside that range there are one real and two complex
states.

The symbols in figure~\ref{fig4} mark the results of the GPE obtained
with a variational approach using six coupled Gaussian functions.
They are in excellent agreement with the results of the matrix model
(\ref{eq:matrixmodel}) for the expectation values
$w=\langle\psi|x^2-y^2|\psi\rangle$ and the matrix model
(\ref{eq:matrixmodel_mf}) for the mean-field energy.
The perfect agreement even for the two parameter perturbations is
remarkable since both matrix models only use two adjustable parameters
to describe the level splittings in equations (\ref{eq:traceless_e})
and (\ref{eq:traceless_w}), which have been determined solely with the
data of the one parameter perturbations.
Similar results as shown in figure~\ref{fig4} for the two parameter
perturbations of the pitchfork bifurcation P2 of the ground state are
also obtained (but not shown) for the pitchfork bifurcation P1 of the
excited state.

\begin{figure}
\includegraphics[width=\textwidth]{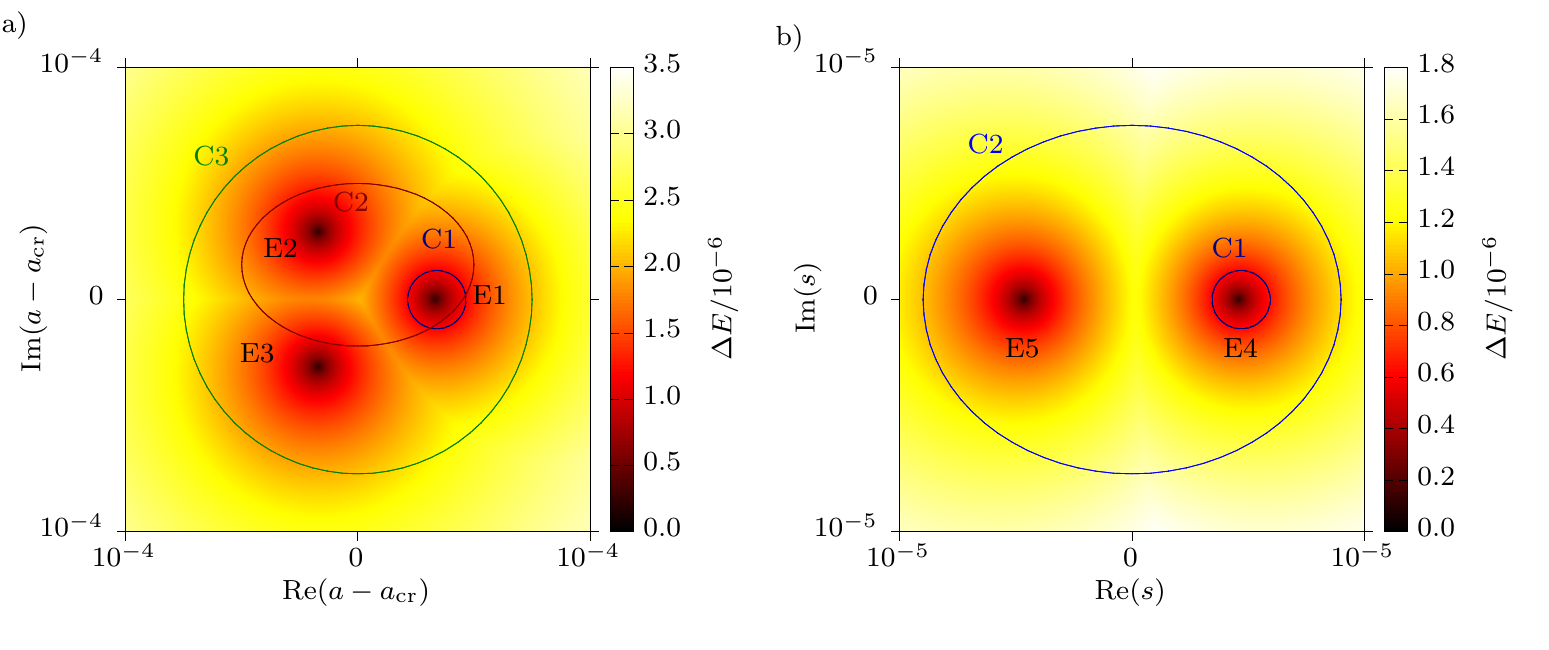}
\includegraphics[width=\textwidth]{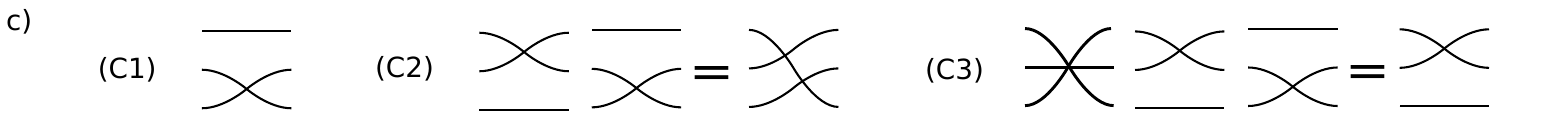}
\caption{Two parameter perturbations of the pitchfork bifurcation P1
  using an ansatz with a single Gaussian function and the trap
  parameters $\bar\gamma_1=34000$ and $\lambda_1=6$. In (a) the
  axial symmetry of the trap is broken with $s=10^{-4}$, and in (b)
  the scattering length is $a-a_{\rm P1}=2\times 10^{-5}$.
  (c) Illustrations of the composed permutations of the states when
  one, two, or three exceptional points are encircled along the paths
  C1 to C3, respectively.}
\label{fig5}
\end{figure}

How do the exceptional points related to the pitchfork bifurcations
behave under the two parameter perturbations?
In order to answer this question we follow different parameter paths
marked C1 to C3 in figure~\ref{fig5} either in the complex scattering
length plane for a constant value $s\ne 0$ (see figure~\ref{fig5}(a))
or in the complex $s$ plane for a constant value $a\ne a_{\rm cr}$
(see figure~\ref{fig5}(b)).
For each path we observe a different permutation behaviour of the
three states, which can be explained as follows.
The colours (or gray values) in figure~\ref{fig5}(a) and (b) indicate
the minimal distance (absolute value) between the mean-field energies
of the three states, i.e.\ a zero value means the coalescence of at
least two states.
In figure~\ref{fig5}(a) three points where two states coalesce are
revealed.
The point marked E1 is a tangent bifurcation point on the real $a$ axis,
as already discussed (see figures~\ref{fig1}(b) and \ref{fig4}(a)).
The points E2 and E3 are new points in the imaginary half planes of
the scattering length where different state pairs coalesce.
If each point is encircled separately they show the permutation
behaviour of an EP2 where the two participating states permute.
Following the path C2 which encircles E1 and E2 the cyclic permutation
of all three states is observed.
If all points E1 to E3 are encircled along the path C3, we find the
same permutation behaviour as for the pitchfork bifurcation of the
dipolar BEC in an axisymmetric trap, i.e.\ two states permute.
The splitting of the EP3 of the pitchfork bifurcation into three EP2
for the broken trap symmetry can only be observed in the permutation
behaviour of the states if a path is chosen which does not include
all three points E1 to E3.
The points E1 to E3 in figure~\ref{fig5}(a) merge at $a=a_{\rm cr}$ in
the limit $s\to 0$ of the asymmetry parameter.

The results in the complex $s$ plane for constant scattering length
$a>a_{\rm cr}$ are presented in figure~\ref{fig5}(b).
Here, the pitchfork bifurcation point P1 is split into the two tangent
bifurcation points E4 and E5.
If the two points are encircled separately each of them shows the
permutation of two states typical of an EP2.
If they are encircled together along the path C2 in
figure~\ref{fig5}(b) the cyclic permutation of all three states as in
figure~\ref{fig3}(b) is observed, i.e.\ the typical behaviour of an EP3.
As in figure~\ref{fig5}(a) the splitting of the EP3 into two
exceptional points with a square root behaviour can only be observed
if a path is chosen which does not include both points E4 and E5, and
these points merge at $s=0$ in the limit $a\to a_{\rm cr}$.

The possible permutations of states when one, two, or three
exceptional points are encircled by a path C1, C2, or C3 is
illustrated in figure~\ref{fig5}(c).
The combinations produce the same permutation behaviour for two and
three EPs as discussed in \cite{Ryu12}.

The exceptional points observed in figure~\ref{fig5}(a) and (b) can
also be obtained with the matrix model in equation
(\ref{eq:matrixmodel}) or (\ref{eq:matrixmodel_mf}).
The eigenvalues of the traceless matrices $Q_w$ or $Q_E$ are the roots
of the characteristic polynomial $\chi(\lambda)=\lambda^3+p\lambda+q$.
Two eigenvalues are degenerate when the discriminant $D=4p^3+27q^2$
vanishes.
This yields the condition
\begin{equation}
  4 \tilde a^3 = 4 c_a^3 (a - a_{\rm cr})^3 = 27 s^2
\label{eq:degeneracy_condition}
\end{equation}
for the scattering length and the asymmetry parameter.
For a constant value $s\ne 0$ equation (\ref{eq:degeneracy_condition})
provides one real and two complex solutions for EP2 exceptional points
in the complex $a$ plane, which agree with the points E1 to E3 marked
in figure~\ref{fig5}(a).
For a given value $a\ne a_{\rm cr}$ of the scattering length equation
(\ref{eq:degeneracy_condition}) provides two real or two imaginary
solutions for $s$, which coincide with the exceptional points E4 and
E5 in figure~\ref{fig5}(b).

\section{Summary}
\label{sec:summary}
When dipolar BECs are described within a mean-field theory the GPE
exhibits a rich variety of nonlinear phenomena including tangent and
pitchfork bifurcations of states and the occurrence of exceptional
points.
We have solved the GPE with an extended variational approach using
coupled Gaussian functions, and presented a method to find all states
participating in the pitchfork bifurcations, including the complex
branches, which have been obtained by analytic continuation of the GPE
using bicomplex numbers.
We have analysed in detail the various bifurcations between the states
depending on two parameters, viz.\ the scattering length $a$ and the
parameter $s$ breaking the axial symmetry of the harmonic trap.
The origin of the bifurcations is the {\em nonlinearity} of the GPE,
nevertheless, the mean-field energies $E_{\rm mf}$ and the expectation
values $w=\langle\psi|x^2-y^2|\psi\rangle$ of the states participating
in the pitchfork bifurcations can be excellently described by a
{\em linear} model with non-Hermitian matrices.

Both the variational computations and the linear model have been used
to investigate the properties of the exceptional points.
At the bifurcation points of a pitchfork bifurcation not only the
three eigenvalues but also the eigenvectors coincide, indicating the
existence of an EP3.
However, a different behaviour for the permutation of states is
observed when the exceptional point is encircled either in the complex
continued scattering length $a$ or in the asymmetry parameter $s$.
The structures resemble those obtained in a linear model using
perturbation theory for non-Hermitian operators \cite{Dem12}.

For two parameter perturbations the pitchfork bifurcation is split into
either three or two tangent bifurcations located in the complex $a$ or
$s$ parameter plane, respectively.
The typical signature of an EP2 is obtained when a single exceptional
point is encircled.
Paths surrounding two or three exceptional points yield the behaviour
of combined permutations \cite{Ryu12}.

The results presented in this article need not be restricted to dipolar
condensates.
Rather, the signatures of exceptional points related to pitchfork
bifurcations discussed here for dipolar condensates may be generic
features of nonlinear systems with pitchfork bifurcations, however,
further investigations will be necessary to clarify this point.
We also expect the results of our investigations to arouse the
interest of experimentalists to search for experimental evidences of
the appearance of exceptional points in Bose-Einstein condensates.

%\section*{Acknowledgments}
\ack
We thank Eva-Maria Graefe for fruitful discussions.
This work was supported by Deutsche Forschungsgemeinschaft.

\appendix

\section{Integrals for the variational ansatz}
\label{sec:app}
The ansatz (\ref{eq:ansatz}) with coupled Gaussians for the wave
functions requires the calculation of Gaussian integrals to set up the
linear system of equations (\ref{eq:linsystem}).
The integrals can be computed analytically or with the help of
elliptic integrals. With the abbreviations
\begin{equation}
  a_\sigma^{kl} = A_\sigma^k + (A_\sigma^l)^* \; , ~
  a_\sigma^{klij} = a_\sigma^{kl} + a_\sigma^{ij} \; ~{\rm with}~~
  \sigma = x,y,z \; ,
\end{equation}
\begin{equation}
  \gamma^{kl} = \gamma^k + (\gamma^l)^* \; , ~
  \gamma^{klij} = \gamma^{kl} + \gamma^{ij} \; , ~
\end{equation}
\begin{equation}
  \kappa_x^{klij} = \sqrt{\frac{a_x^{klij} a_z^{ij} a_z^{kl}}{a_x^{ij} a_x^{kl} a_z^{klij}}} \; , ~
  \kappa_y^{klij} = \sqrt{\frac{a_y^{klij} a_z^{ij} a_z^{kl}}{a_y^{ij} a_y^{kl} a_z^{klij}}}
\end{equation}
the integrals read
\begin{eqnarray}
\fl
  \langle g^l | g^k\rangle &= \frac{\pi^{3/2} \ee^{-\gamma^{kl}}}
    {\sqrt{a_x^{kl} a_y^{kl} a_z^{kl}}} \; , \\
\fl
  \langle g^l | x^2 | g^k\rangle &= \frac{\pi^{3/2} \ee^{-\gamma^{kl}}}
    {2 \left(a_x^{kl}\right)^{3/2}\sqrt{a_y^{kl} a_z^{kl}}} \; , \\
\fl
  \langle g^l | x^4 | g^k\rangle &= \frac{3\pi^{3/2} \ee^{-\gamma^{kl}}}
    {4 \left(a_x^{kl}\right)^{5/2}\sqrt{a_y^{kl} a_z^{kl}}} \; , \\
\fl
 \langle g^l | x^2y^2 | g^k\rangle &= \frac{\pi^{3/2} \ee^{-\gamma^{kl}}}
   {4\left(a_x^{kl} a_y^{kl}\right)^{3/2} \sqrt{a_z^{kl}}} \; , \\
\fl
 \langle g^l | V_{\rm c} | g^k\rangle &=
   8 a \pi^{5/2} \sum_{i,j=1}^N \frac{\ee^{-\gamma^{klij}}}{
     \sqrt{a_x^{klij} a_y^{klij} a_z^{klij}} } \; , \\
\fl
 \langle g^l | x^2V_{\rm c} | g^k\rangle &=
   4 a \pi^{5/2} \sum_{i,j=1}^N \frac{\ee^{-\gamma^{klij}}}{
     \left(a_x^{klij}\right)^{3/2} \sqrt{a_y^{klij} a_z^{klij}} } \; , \\
%\end{eqnarray}
%\begin{eqnarray}
\label{eq:1vd}
\fl
  \langle g^l | V_{\rm d} | g^k\rangle &=
    \frac{4 \pi^{5/2}}{3} \sum_{i,j=1}^N \frac{\ee^{-\gamma^{klij}}}
    {\sqrt{a_x^{klij}a_y^{klij}a_z^{klij}}} \left[ \kappa_x \kappa_y
      R_D\left( \kappa_x^2, \kappa_y^2, 1\right) - 1\right] \; , \\
\label{eq:x2vd}
\fl
  \langle g^l | x^2V_{\rm d} | g^k\rangle &=
    \frac{4 \pi^{5/2}}{3} \sum_{i,j=1}^N \frac{\ee^{-\gamma^{klij}}}
    {\sqrt{a_x^{klij}a_y^{klij}a_z^{klij}}}
%    \nonumber\\ &\phantom{=\;} \times
 \Bigg[ 
       \frac{1}{2 a_x^{klij}}  \left( \kappa_x \kappa_y R_D\left( \kappa_x^2, \kappa_y^2, 1\right) - 1 \right)
      \nonumber\\&\phantom{=\;}- \bigg(
          \left(
              \kappa_y R_D\left( \kappa_x^2, \kappa_y^2, 1\right)
            + 2 \kappa_x^2 \kappa_y R_x\left( \kappa_x^2, \kappa_y^2, 1\right)
          \right) \frac{\partial \kappa_x}{\partial A_x^k} \nonumber\\ &\phantom{=\;} -
          \left(
              \kappa_x R_D\left( \kappa_x^2, \kappa_y^2, 1\right)
            + 2 \kappa_x \kappa_y^2 R_y\left( \kappa_x^2, \kappa_y^2, 1\right)
          \right) \frac{\partial \kappa_y}{\partial A_x^k}
        \bigg)
    \Bigg] \; , \\
\label{eq:y2vd}
\fl
 \langle g^l | y^2V_{\rm d} | g^k\rangle &=
  \frac{4 \pi^{5/2}}{3} \sum_{i,j=1}^N \frac{\ee^{-\gamma^{klij}}}{\sqrt{a_x^{klij}a_y^{klij}a_z^{klij}}}
%    \nonumber\\ &\phantom{=\;}    \times
 \Bigg[ 
       \frac{1}{2 a_y^{klij}}  \left( \kappa_x \kappa_y R_D\left( \kappa_x^2, \kappa_y^2, 1\right) - 1 \right)
      \nonumber\\ &\phantom{=\;}- \bigg(
          \left(
              \kappa_y R_D\left( \kappa_x^2, \kappa_y^2, 1\right)
            + 2 \kappa_x^2 \kappa_y R_x\left( \kappa_x^2, \kappa_y^2, 1\right)
          \right) \frac{\partial \kappa_x}{\partial A_y^k} \nonumber\\ &\phantom{=\;} -
          \left(
              \kappa_x R_D\left( \kappa_x^2, \kappa_y^2, 1\right)
            + 2 \kappa_x \kappa_y^2 R_y\left( \kappa_x^2, \kappa_y^2, 1\right)
          \right) \frac{\partial \kappa_y}{\partial A_y^k}
        \bigg)
    \Bigg] \; , \\
\label{eq:z2vd}
\fl
 \langle g^l | z^2V_{\rm d} | g^k\rangle &=
 \frac{4 \pi^{5/2}}{3} \sum_{i,j=1}^N \frac{\ee^{-\gamma^{klij}}}{\sqrt{a_x^{klij}a_y^{klij}a_z^{klij}}}
%    \nonumber\\ &\phantom{=\;}    \times
 \Bigg[ 
      \frac{1}{2 a_z^{klij}}  \left( \kappa_x \kappa_y R_D\left( \kappa_x^2, \kappa_y^2, 1\right) - 1 \right)
      \nonumber\\ &\phantom{=\;}- \bigg(
          \left(
              \kappa_y R_D\left( \kappa_x^2, \kappa_y^2, 1\right)
            + 2 \kappa_x^2 \kappa_y R_x\left( \kappa_x^2, \kappa_y^2, 1\right)
          \right) \frac{\partial \kappa_x}{\partial A_z^k} \nonumber\\ &\phantom{=\;} -
          \left(
              \kappa_x R_D\left( \kappa_x^2, \kappa_y^2, 1\right)
            + 2 \kappa_x \kappa_y^2 R_y\left( \kappa_x^2, \kappa_y^2, 1\right)
          \right) \frac{\partial \kappa_y}{\partial A_z^k}
        \bigg)
    \Bigg] \; .
\end{eqnarray}
In equations (\ref{eq:1vd})-(\ref{eq:z2vd}) the upper indices at
$\kappa_x^{klij}$ and $\kappa_y^{klij}$ have been omitted.
The elliptic integrals $R_D(x,y,z)$ are given in equation
(\ref{eq:elliptic_integral}), and their derivatives are defined as
\begin{eqnarray}
  R_x(x,y,z) = \partial_x R_D(x,y,z) \; , \;
  R_y(x,y,z) = \partial_y R_D(x,y,z) \; .
\end{eqnarray}
Integrals of the harmonic trap $V_{\rm t}$ are easily obtained with
the equations given above.
Integrals not given above are obtained by appropriate permutations of
$x$, $y$ and $z$.

\section*{References}
%
%\bibliographystyle{unsrt}
%\bibliography{paper}

\end{document}